\documentclass[a4paper,10pt,twocolumn]{article}

\usepackage[english]{babel}
\usepackage[utf8]{inputenc}
\usepackage[T1]{fontenc}

\usepackage[top=1.5cm, left=1.5cm, right=1.5cm, bottom=1.5cm]{geometry}

\renewenvironment{abstract}{\bf\small {\em\ Abstract---}}{}

\usepackage{amsfonts,amssymb,amsmath,amsthm}
\usepackage{subfigure}
\usepackage{graphicx}
\usepackage[footnotesize]{caption}

\usepackage{mathrsfs}
\usepackage{algorithm}
\usepackage{algorithmic}
\usepackage{xcolor}
\usepackage{enumitem}

\title{Joint nonstationary blind source separation and spectral analysis}

\author{Adrien Meynard$^1$\\
  \footnotesize $^1$Aix Marseille Univ, CNRS, Centrale Marseille, I2M, Marseille, France.} \date{\empty} 
  
\def\bz{{\mathbf z}}
\def\bA{{\mathbf A}}
\def\bB{{\mathbf B}}

\def\bM{{\mathbf M}}
\def\by{{\mathbf y}}
\def\bs{{\mathbf s}}
\def\bw{{\mathbf w}}

\def\ccS{\mathscr{S}}
\def\cD{\mathcal{D}}
\def\cW{\mathcal{W}}
\def\RR{\mathbb{R}}

\def\btheta{\boldsymbol{\theta}}
\def\bSigma{\boldsymbol{\Sigma}}

\def\defeq{\stackrel{\Delta}{=}}

\begin{document}

\maketitle

\begin{abstract} 
We address a nonstationary blind source separation (BSS) problem. The model includes both nonstationary sources and mixing. Therefore, we introduce an algorithm for joint BSS and estimation of stationarity-breaking deformations and spectra. Finally, its performances are evaluated on a synthetic example. 
\end{abstract}

\section{Introduction: model and background}
The BSS problem, originally introduced in a stationary context, has also been discussed in nonstationary situations. Extensions to nonstationary signals have been proposed, based on time-frequency analysis (see~\cite{Belouchrani13source}, chap.~9 in~\cite{Jutten07separation} and references therein), or based on mutual information~\cite{Pham01blind}. The BSS of a nonstationary mixtures of stationary signals have also been studied. For instance, in~\cite{Parra00convolutive}, the authors explore the convolutive BSS problem. In the following, we tackle a doubly nonstationary BSS problem, and propose a demultiplexing algorithm adapted to a specific class of nonstationary signals mixed by a instantaneous nonstationary mixing matrix.

\subsection{Nonstationarity}
The nonstationary signals of interest here are deformed versions of stationary signals. 

Let $x$ denote a stationary signal, modeled as a realization of a stationary random process with power spectrum denoted by $\ccS_X$. Acting on $x$ with a stationarity-breaking operator yields a nonstationary signal denoted by $y$. Various classes of stationarity-breaking operators are relevant to model physical phenomena (e.g.~frequency modulation~\cite{Meynard17spectral}, amplitude modulation~\cite{Meynard18spectral}). We focus here on the time warping operator denoted by $\cD_\gamma$ and defined by:
\begin{equation}
\label{eq:nonstat.model}
y(t) = \cD_\gamma x (t)= \sqrt{\gamma'(t)}\,x(\gamma(t))\ ,
\end{equation}
where $\gamma\in C^2$ is a strictly increasing smooth function. Such deformations can model  nonstationary physical phenomena as diverse as Doppler effect, speed variations of an engine, animal vocalization or speech~\cite{Stowell18computational,Meynard18spectral}.

The wavelet transform is a natural tool to analyze such signals. Hence, the wavelet transform $\cW_x$ of the signal $x$ is defined by:
\begin{equation}
\cW_x(s,\tau) = \int_\RR x(t)q^{-s/2}\overline{\psi}\left(\dfrac{t-\tau}{q^s}\right)dt\quad \text{with}\quad q>1\ .
\end{equation}
In that framework, it can be shown that the respective wavelet transforms $\cW_y$ and $\cW_x$ of $y$ and  $x$ are approximately related by
\begin{equation}
\label{eq:approx.wavelet}
\cW_y(s,\tau) \approx \cW_x(s + \log_q(\gamma'(\tau)),\gamma(\tau))\ .
\end{equation}
In the following, we make the assumption that $x$ is a realization of a stationary random process $X$. In such a setting, the approximation error can be controlled thanks to the decay properties of the wavelet $\psi$, and the variations of $\gamma'$. In~\cite{Meynard17spectral,Meynard18spectral}, corresponding quantitative error bounds are given.

\subsection{Blind source separation}
\vspace{-2mm}

The problem we consider is the BSS of nonstationary signals modeled by equation~\eqref{eq:nonstat.model}.

We investigate the case where the number of sources and the number of observations are equal and denoted by $N$. The sources are additionally assumed to be independent. Let $\by(t),\bz(t)\in\RR^N$ denote the column vectors containing respectively all the sources and observations at time $t$. Then, the mixture is written as
\vspace{-1mm}
\begin{equation}
\label{eq:bss.model}
\bz(t) = \bA(t)\by(t)\ ,
\end{equation}
where $\bA(t)\in\RR^{N \times N}$ denotes the time varying mixing matrix, assumed to be invertible. This model generalizes the amplitude modulation model in the case $N=1$ detailed in~\cite{Meynard18spectral}. For example, this model can be appropriate in bioacoustics to describe the BSS of a howling wolf pack~\cite{Passilongo15vizualizing,Papin18acoustic}. 

Our goal is to determine jointly the mixing matrix $\bA(t)$, the time warping functions $\gamma_i(t)$, and the spectra of the stationary sources $\ccS_{X_i}$ for $i=1,\ldots,N$ from the observations $\bz(t)$.

Let us consider a fixed time $\tau$, then for each observation $z_i$, we denote by $\bw_{z_i,\tau} = \cW_{z_i}(\bs,\tau)$ the row vector containing the values of the wavelet transform for a vector of scales  $\bs$ (of size denoted by $M_s$). Then, all these vectors are gathered into a $N \times M_s$ matrix $\bw_{\bz,\tau}$ such that $\bw_{\bz,\tau} = \left(\bw_{z_1,\tau}^T \cdots \bw_{z_N,\tau}^T \right)^T$. The same operation is applied to the wavelet transform of the sources. The matrix $\bA(t)$ is assumed to vary slowly with respect to the oscillations of the signals. It can be shown that the linear relation~\eqref{eq:bss.model} becomes in this new setting a relationship between the wavelet transforms of $\by$ and $\bz$ of the form
\vspace{-1mm}
\begin{equation}
\label{eq:approx.bss}
\bw_{\bz,\tau} \approx \bA(\tau)\bw_{\by,\tau}\ .
\end{equation}
Aside from the terms controlling the error bound in~\eqref{eq:approx.wavelet}, the error bound in~\eqref{eq:approx.bss} is also controlled by the variations of the mixing matrix coefficients.

\vspace{-3mm}
\section{Estimation procedure}
\vspace{-2mm}
Approximation equations~\eqref{eq:approx.wavelet} and~\eqref{eq:approx.bss} allow us to write an approximate likelihood in the Gaussian case (see~\cite{Cardoso98blind} for more details on this approach).

The estimation procedure is based upon discrete wavelet transforms, time-varying parameters are therefore estimated on a discrete time grid $D$. In the following, the estimation procedure is described for a given $\tau\in D$.
For the sake of simplicity, we introduce the following notations: $\bB_\tau = \bA(\tau)^{-1}$, $\theta_{i,\tau} = \log_q\left(\gamma'_i(\tau)\right)$ and $\btheta_\tau = (\theta_{1,\tau} \cdots \theta_{N,\tau})^T$.

\vspace{-2mm}
\subsection{Probabilistic setting}
\vspace{-2mm}

It follows from the Gaussianity assumption on $X$ that $\bw_{y_i,\tau}\sim \mathcal{N}(\mathbf{0},\bSigma_i(\theta_{i,\tau}))$, where
\[
\left[\bSigma_i(\theta_{i,\tau})\right]_{kk'} = q^{\frac{s_k+s_{k'}}{2}}\!\!\int_\RR\! \ccS_{X_i}(q^{-\theta_{i,\tau}}\xi)\overline{\hat\psi}\left(q^{s_k}\xi\right)\hat\psi\left(q^{s_{k'}}\xi\right) d\xi.
\]
Let $p_V$ denote generically the probability density function of a random vector $V$. Then, the source independence hypothesis gives the following opposite of the log-likelihood:\\[-5mm]
\begin{align*}
\ell_\tau(\bB_\tau,\btheta_\tau) \defeq& -\log(p_{\bw_{z,\tau}|(\bB_\tau,\btheta_\tau)}(\bw_{\bz,\tau};\bB_\tau,\btheta_\tau)) + c \\[-2mm]
 =& -M_s \log|\det(\bB_\tau)| \!+\!\dfrac1{2}\sum_{i=1}^N\log\left|\det\bSigma_i(\theta_{i,\tau}))\right| \\[-3mm]
&+\dfrac1{2}\sum_{i=1}^N[\bB_\tau\bw_{\bz,\tau}]_{i\cdot}\bSigma_i(\theta_{i,\tau})^{-1}[\bB_\tau\bw_{\bz,\tau}]_{i\cdot}^H\ ,
\end{align*}\\[-3mm]
where $[\bM]_{i\cdot}$ denotes the $i$-th line of the matrix $\bM$, and $\bM^H$ is its conjugate transpose. Maximum likelihood (ML) estimates, i.e. minimizers of $\ell_\tau(\bB_\tau,\btheta_\tau)$, can be evaluated numerically.

However, in order to take into account the smoothness assumption on the mixing matrix with respect to time, we switch to the Bayesian framework and introduce a prior $p_{\bB_\tau}$ on the unmixing matrix $\bB_\tau$ (assuming i.i.d. matrix coefficients). We choose for $p_{\bB_\tau}$ a uniform distribution centered on $\bB_{\tau-\Delta_\tau}$, and with support $2\epsilon_B\Delta_\tau$. Then, the maximum a posteriori (MAP) estimate $\tilde\bB_\tau$ can be written as the solution of the problem
\begin{equation}
\tilde\bB_\tau = \arg\min_{\bB_\tau}\ \ell_\tau(\bB_\tau,\btheta_\tau)\quad\text{s.t.}\quad\|\bB_{\tau} - \bB_{\tau-\Delta_\tau}\|_\infty\leq\epsilon_B\Delta_\tau.
\label{eq:unif.pb}
\end{equation}
This problem is consistent with the smoothness hypothesis on $\bB_\tau$. Indeed, assuming $\Delta_\tau$ is small, the constraint in equation~\eqref{eq:unif.pb} is almost equivalent to $\|\bB_{\tau}'\|_\infty \leq \epsilon_B$.

Concerning the time warping estimation, we choose not to give a prior on $\btheta_\tau$. Thus, $\tilde\btheta_\tau$ is the ML estimation of $\btheta_\tau$. 

\vspace{-2mm}
\subsection{Estimation algorithm}
\vspace{-1mm}
The estimation strategy is to alternate the estimations of $\bB_\tau$, $\btheta_\tau$ and the spectra. The algorithm~\ref{alg:estimation} (named JEFAS-BSS) synthesizes all the estimation steps which are described below.
\vspace{-1mm}
\begin{itemize}[leftmargin=*]
\item\textit{Mixing matrix estimation.}
In practice, we numerically solve the problem~\eqref{eq:unif.pb}. Besides, because of the assumption of slow variations of the matrix coefficients, we make the approximation that $\bB_\tau$ is constant on the interval $I_\tau = [\tau-\Delta_\tau/2,\ \tau+\Delta_\tau/2[$. Finally, the estimated sources $\tilde\by_\tau$ are obtained via $\tilde\by_\tau(t) = \tilde\bB_\tau\bz(t)$ where $t\in I_\tau$.
Notice that for each interval $I_\tau$, a new matrix $\tilde\bB_\tau$ is applied to the observations. Due to the source ordering indeterminacy, a reordering method has to be introduced to connect consecutive segments of each source signal. We use for that the Gale-Shapley stable marriage algorithm~\cite{Gale62college} which constructs stable matchings between consecutive time slices source estimations. The ranking criterion is based on the comparison of the dot products between normalized Fourier spectra of these slices.
\vspace{-1mm}
\item\textit{Deformations and spectra estimations.}
For each source, the joint estimation of $\left\lbrace\theta_{i,\tau}\right\rbrace_{\tau\in D}$ and $\ccS_{X_i}$ is obtained via the JEFAS algorithm (which is detailed in~\cite{Meynard18spectral}). For this purpose, the input wavelet transform $\bw_{\by}$ of the source $y_i$ is replaced with its estimate $\left\lbrace\bB_\tau\bw_{\bz,\tau}\right\rbrace_{\tau\in D}$.

\end{itemize}

\vspace{-1mm}
Regarding initialization, a basic method is to use a stationary BSS method on observations to obtain a first unmixing matrix estimate. For instance, SOBI~\cite{Belouchrani97blind} is a stationary BSS algorithm which can give an initial unmixing matrix. A better initial matrix can be obtained by piecewise SOBI estimates on non overlapping segments (called p-SOBI), where the stationarity assumption makes more sense.

\setlength{\textfloatsep}{10pt}
\begin{algorithm}[t]
\caption{JEFAS-BSS}
\label{alg:estimation}
\begin{algorithmic}

\STATE {\bf Initialization:} Obtain $\tilde\bB^{(0)}_\tau$ by means of the p-SOBI algorithm. Compute the estimated source $\tilde\by^{(0)}(\tau) = \tilde\bB^{(0)}_\tau\bz(\tau)$.

\STATE $\bullet$ $k \leftarrow 1$

\WHILE{stopping criterion is false \AND $k\leq k_{max}$}

\STATE $\bullet$ For $i=1,\ldots,N$, estimate parameters $\tilde\theta^{(k)}_{i,\tau},\ \forall\tau\in D$ and spectrum $\tilde\ccS_{X_i}^{(k)}$ applying JEFAS algorithm to $\tilde y_i^{(k-1)}$.

\FOR{$\tau=0,\Delta_\tau,\ldots,T$}

\STATE $\bullet$ Estimate $\tilde\bB^{(k)}_\tau$: solve~\eqref{eq:unif.pb} replacing $\btheta_\tau$ and $\ccS_X$ with their current estimations $\tilde\btheta_\tau^{(k)}$ and $\left\lbrace\tilde\ccS_{X_i}^{(k)}\right\rbrace_{i=1,\ldots,N}$.

\ENDFOR

\STATE $\bullet$ Estimate the sources $\tilde\by^{(k)}$.

\STATE $\bullet$ $k\leftarrow k+1$

\ENDWHILE
\end{algorithmic}
\end{algorithm}

The convergence is monitored using the Source to Interference Ratio (SIR) introduced in~\cite{Vincent06performance}. For a given estimated source, SIR quantifies the presence of interferences from the other true sources. As we do not have access to the ground truth sources, we use as stopping criterion the SIR between $\tilde \by^{(k-1)}$ and $\tilde \by^{(k)}$ (instead of $\by$) which gives an evaluation of the BSS update, and is therefore a relevant convergence assessment.

\section{Results}
\vspace{-3mm}

\begin{figure}[h]
\vspace{-3mm}
\centering
\includegraphics[width=0.24\textwidth]{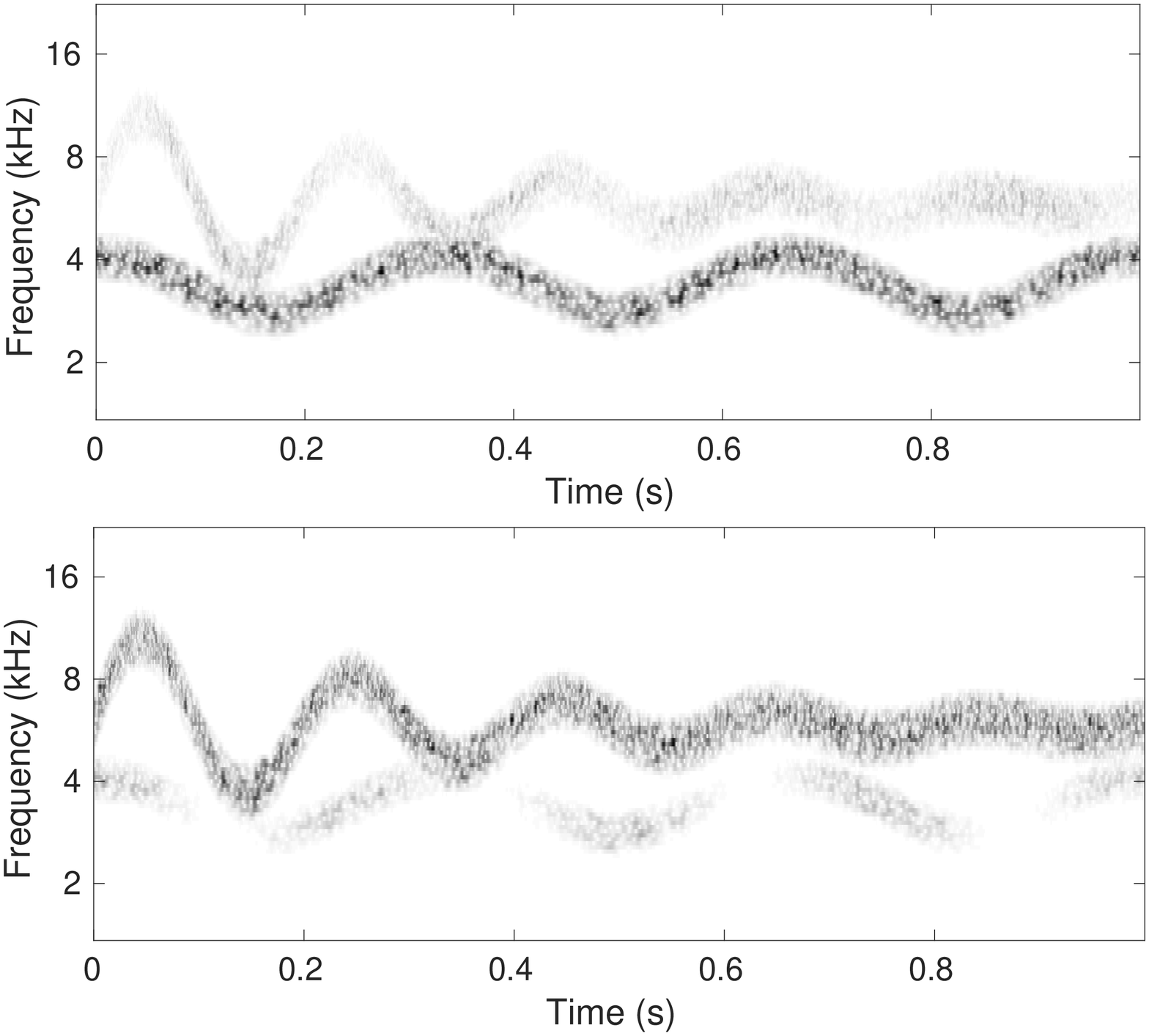}
\includegraphics[width=0.24\textwidth]{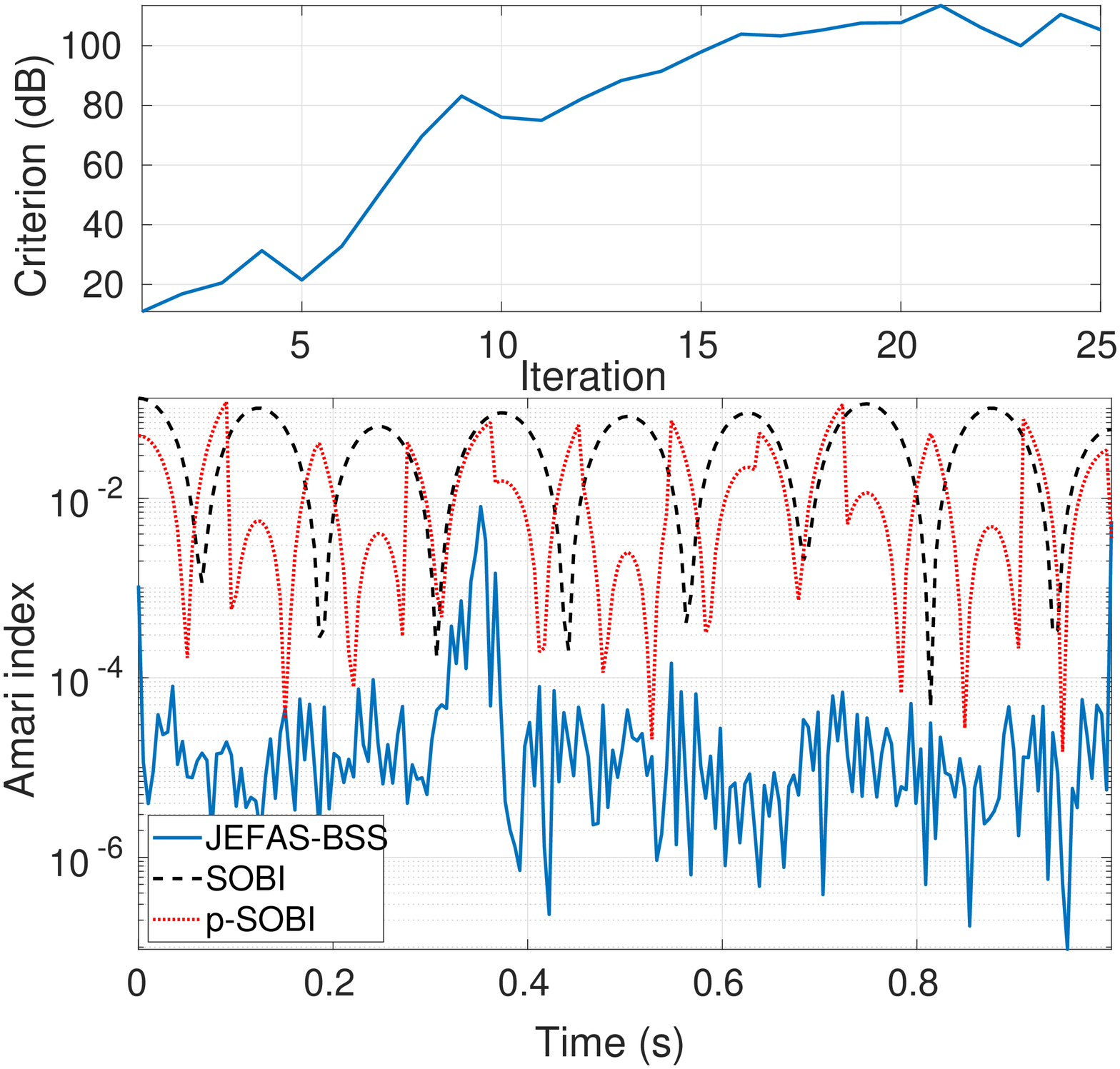}

\vspace{-4mm}
\caption{Synthetic example. Left: scalograms of the observations. Top-right: Convergence criterion evolution. Bottom-right: Amari index evolution.}
\label{fig:bss.ns}
\vspace{-6mm}
\end{figure}

We construct a synthetic example to evaluate the performances of the algorithm. The two sources are band-pass filtered white noise, with time-varying bandwidth. The mixing matrix coefficients are sinusoidally varying over time. On the left of figure~\ref{fig:bss.ns}, the wavelet transforms of both observations are displayed.

The evolution of the convergence criterion through iterations of JEFAS-BSS is displayed in figure~\ref{fig:bss.ns} (top-right). We can empirically note that our algorithm converges in a small number of iterations. Indeed, after 15~iterations the convergence criterion is around 100~dB meaning the BSS update is negligible. 

Finally, we evaluate the performances of the BSS algorithms (we refer to~\cite{Meynard18spectral} for the evaluation of the performances of the deformations and spectra estimations). The Amari index~\cite{Amari96new} is a measure of divergence between the matrix $\tilde\bB_\tau \bA_\tau$ and the identity matrix. The closer to zero the Amari index the better. On the bottom-right of figure~\ref{fig:bss.ns}, we display the evolution of the Amari index through time for each BSS algorithm. In table~\ref{tab:perf}, we also compare the SIR, the SDR (Source to Distortion Ratio~\cite{Vincent06performance}), and the time-averaged Amari index of the BSS algorithms. Those different criteria show that BSS-JEFAS performances are higher than those of SOBI and p-SOBI. Besides, in average, p-SOBI gives a better Amari index than SOBI, which is understandable because it takes into account the nonstationarity of the mixing matrix. Nonetheless, the SIR and SDR of p-SOBI are worse than those of SOBI. Indeed, because this method does not take into account the regularity of $\bB_\tau$, the connections between slices are sensitive to discontinuities and create distortion in the estimated sources.

\begin{table}[h]
\vspace{-1mm}
\begin{tabular}{|l||c|c|c|}
  \hline
   Criterion & SOBI & p-SOBI & JEFAS-BSS\\
   \hline 
   \hline
   SIR (dB) & $28.55$ & $15.04$ & $46.55$ \\
   SDR (dB) & $16.60$ & $-4.53$ & $37.69$ \\
   Amari index & $4.63\times 10^{-2}$ & $1.74\times\! 10^{-2}$ & $1.40\times 10^{-4}$ \\
   \hline
\end{tabular}

\vspace{-3mm}
\caption{Comparison of the performances between BSS algorithms: standard SOBI, piecewise SOBI and the proposed algorithm.}
\label{tab:perf}
\end{table}

\bibliographystyle{ieeetr}
\bibliography{Sampta17}

\end{document}